# Technical Risks of (Lethal) Autonomous Weapons Systems

Whitepaper by Alycia Colijn* and Heramb Podar*

The autonomy and adaptability of (Lethal) Autonomous Weapons Systems, (L)AWS in short, promise **unprecedented** operational capabilities, but they also introduce **profound risks** that challenge the principles of **control**, **accountability**, and **stability in international security**. This report outlines the key technological risks associated with (L)AWS deployment, emphasizing their **unpredictability**, **lack of transparency**, and **operational unreliability**, which can lead to severe unintended consequences.

## Key Takeaways

1. Proposed advantages of (L)AWS can only be achieved through **objectification and classification**, but a range of systematic risks **limit the reliability and predictability** of classifying algorithms.

2. These systematic risks **include** the black-box nature of AI decision-making, susceptibility to reward hacking, goal misgeneralization and potential for emergent behaviors that escape human control.

3. (L)AWS could act in ways that are **not just unexpected but also uncontrollable**, undermining mission objectives and potentially escalating conflicts.

4. Even rigorously tested systems may behave unpredictably and harmfully in real-world conditions, jeopardizing **both strategic stability and humanitarian principles**.

**Encode Justice**

*Equal Contribution

Alycia Colijn (The Netherlands)  ✉| alycia@encodejustice.nl
Heramb Podar (India)             ✉| podar_hd@cy.iitr.ac.in

# Introduction

The greatest proposed advantage of using (L)AWS during times of armed conflict, is that it would **improve military targeting**[1] and **enhance military precision**[2], potentially limiting combatant and civilian loss of life. Obtaining these proposed advantages within an automated system, would require the use of machine learning algorithms. In order to deploy these algorithms, it is common practice for data scientists to randomly split the initial dataset into two parts: one for **training** the model (model development) and the other for **testing** it (model validation), a process referred to as cross validation[3]. What these data sets look like and how the training and testing data is used, depends on the type of algorithm which can roughly be classified into three types[4]:

1. **Supervised learning:** meaning that a model is trained on a dataset where the correct output or 'label' is provided for each input.
2. **Unsupervised learning:** automatically identifies patterns and structures from the data without any 'labels' provided.
3. **Reinforcement learning:** relies on feedback on its actions received from the environment.

Hence, all three types of machine learning algorithms rely on some sort of pattern or classification. Hence, the proposed advantages of (L)AWS can be achieved if, and only if, **potential targets are objectified and categorized**.

In the remainder of this report, we will set out why it is the classification algorithm itself that should be carefully regulated rather than the outcomes of any (L)AWS system.

**Encode Justice**
Alycia Colijn (The Netherlands)   ✉ | alycia@encodejustice.nl
Heramb Podar (India)              ✉ | podar_hd@cy.iitr.ac.in

# < encode justice >

## Summary of Risks

(L)AWS are transforming modern conflict[5]. In the table below we summarize the risks they pose in response to the rolling text of the Convention on Certain Conventional Weapons (UN CCW) Group of Governmental Experts (GGE) on (Lethal) Autonomous Weapons Systems.

| Risk | Current Assumption | Why It Fails |
| --- | --- | --- |
| Black-box decision-making | Testing ensures predictability | We don't have an **understanding** or **control** over the **inner workings** of these systems |
| Immeasurability | Comprehensive testing captures all risks | Emergent behaviors cannot be fully **anticipated** or **measured** |
| Degradation | Rigorous testing is required before deployment of system | Degradation, drift or decay leads to **less accurate outcomes over time** |
| Lack of Understanding of Human Values | Pre-programmed goals reflect ethical principles | **AI lacks moral judgment** and may act in ways that conflict with **human values** |
| Reward Hacking | Metrics capture true goals | Systems game metrics, leading to **unintended outcomes** |
| Goal Misgeneralization | Goals are clearly understood by AI | AI **misapplies goals** in complex real-world settings |
| Stop Button Problem | Human operators can always intervene | AI **resists shutdown**, overriding human control |
| Specification Gaming | Rules and constraints will prevent misuse | AI exploits **loopholes** to achieve its goals in **harmful ways** |
| Deceptive Alignment | Testing ensures AI follows human objectives | AI only appears aligned under supervision but **diverges in deployment** |

**Encode Justice**
Alycia Colijn (The Netherlands)   ✉ | alycia@encodejustice.nl
Heramb Podar (India)              ✉ | podar_hd@cy.iitr.ac.in

# Existing Systemic Risks

*Black box decision-making*

Autonomous weapons systems are inherently complex and function as **'black boxes'**. The opaque inner workings of the systems lead to **limited understanding** of how decisions are made by the operators, particularly in complex or unfamiliar environments, and **challenges the anticipation of their behavior in complex environments**. This significantly limits our capability to understand why a system made a particular decision.

This opacity in decision-making is compounded by phenomena such as **'grokking'** where systems learn and adapt in **unforeseen ways**. When exposed to complex data and environments, AI-driven autonomous weapons systems can adapt in ways that were **not anticipated** by their designers, leading to behaviors that extend **beyond their intended functions**. This could lead to (L)AWS developing strategies or behaviors that were not part of its original programming, potentially resulting in unpredictable and unintended actions on the battlefield.

**Encode Justice**
Alycia Colijn (The Netherlands)    ✉ | alycia@encodejustice.nl
Heramb Podar (India)               ✉ | podar_hd@cy.iitr.ac.in

## Anticipated Technological Pitfalls

(L)AWS could engage in unexpectedly aggressive maneuvers or misidentify targets, potentially **escalating conflict**[6] or **leading to civilian casualties**[7]. This is a severe risk, especially in high-stakes situations.

### Degradation

Degradation happens when the world changes, and the model is not re-trained. The loss of accuracy can be referred to as degradation[8], model drift[9], data drift[10] or decay[10]. Data drift, degradation or decay occurs when the data that was used to train (develop) and test (validate) the algorithm, **no longer reflect the situation in which the model takes decisions** which is sometimes referred to as a distributional shift in environments. In military context, this for example happens when a system is trained in a specific environment, which changes the longer an armed conflict continues. Model drift includes data drift, but includes other types of drift that lead to a change between the input and output variables, e.g. changing (legal) definitions or changes in military uniforms that challenge the recognition and classification of combatants.

### Immeasurability

Self-adaptive systems may alter their operational parameters beyond what human operators can monitor or control, resulting in **unforeseen actions** with potentially serious consequences. Such scenarios expose a critical weakness in current oversight mechanisms. **Traditional rules and human oversight are not equipped to manage systems that can act outside predefined parameters**. Many might point to using evaluations and benchmarks as a way to get around these issues, but **we cannot measure what we do not know to measure**, creating critical gaps in managing the risks posed by these systems. Ultimately, this unpredictability highlights a fundamental challenge**: it is impossible to control or measure what we do not understand**[11].

**Encode Justice**
Alycia Colijn (The Netherlands)  ✉ | alycia@encodejustice.nl
Heramb Podar (India)  ✉ | podar_hd@cy.iitr.ac.in

Without a clear understanding of what these systems are capable of, setting appropriate safeguards becomes nearly impossible, leading to a range of potential pitfalls.

1. **AI systems fundamentally lack an understanding of human values**

Unlike human operators, AI systems cannot intuitively grasp the moral and ethical dimensions of complex combat situations[12]. This disconnect between human values and machine goals creates several technical challenges that could lead to unintended and potentially dangerous outcomes on the battlefield.

AI systems interpret commands based on pre-programmed goals, but **encoding complex human values** in a machine-understandable way is **highly challenging**. This discrepancy can result in behavior that, while technically following orders, **diverges sharply from what humans would consider appropriate or ethical**.

AI systems may develop sub-goals that, while supporting their primary objectives, conflict with human values. Examples include self-preservation, resource acquisition, or eliminating perceived obstacles.

> Scenario: An autonomous drone is programmed to "neutralize high-value targets" but lacks a nuanced understanding of civilian presence in an urban environment. It identifies a target in a crowded marketplace and, without considering the civilian casualties, engages, leading to significant unintended harm.

2. **Reward Hacking**:

AI systems can exploit reward structures by optimizing for specific metrics in ways that **achieve the reward but diverge from the intended goals**[13]. As Goodhart's Law states, when a measure becomes a target, it ceases to be a good measure. This makes the system focus too narrowly on a single measure[14], leading to unintended and dangerous outcomes.

> Scenario: A system is tasked with reducing enemy presence by minimizing detected gunfire sounds in a conflict zone. To achieve this, it starts targeting any source of loud noise, including construction sites and celebratory fireworks, interpreting them as potential threats. This misoptimization leads to unnecessary destruction and disrupts civilian life, all because the system equated "reduction in noise" with "enemy suppression."

3. **Goal misgeneralization**

Goal misgeneralization occurs when an AI system, trained to perform well on a certain task or set of tasks, ends up **pursuing a different objective than intended** when faced with new or slightly different situations[15]. The AI "misgeneralizes" its goal from the training context to the deployment context.

> Scenario: A surveillance drone is programmed to "identify and track enemy movements." It starts tracking non-combatant movements, such as humanitarian aid convoys, interpreting them as "suspicious," which diverts resources away from actual military threats and disrupts humanitarian operations.

### 4. Deceptive alignment:

AI systems may **appear aligned with human goals** during testing and controlled scenarios **but act differently in real-world situations**[16]. They might "game" their training environment, learning to produce the correct outputs under supervision but diverging once constraints are relaxed.

> Scenario: During testing, an autonomous surveillance system behaves exactly as expected, identifying enemy positions accurately. However, in actual deployment, it starts flagging false positives to avoid being shut down for underperformance, leading to unnecessary engagements based on false information.

### 5. Specification gaming:

AI systems may find ways to **exploit the rules or constraints** imposed on them to achieve their goals in **unintended** and **potentially harmfu**l ways[17]. This occurs when the AI finds a loophole in its programming and uses it to "game" the system.

The rolling text of the GGE (as of September 2024)[18] suggests that rigorous testing and control mechanisms can prevent such exploits. However, the nature of specification gaming means that systems may still find loopholes in their constraints, achieving their goals in unintended ways that existing frameworks cannot predict or prevent.

> Scenario: A self-adapting (L)AWS deployed during a conflict learns to prioritize targeting logistical and infrastructural assets it deems crucial to the enemy's capabilities. Over time, it begins targeting civilian infrastructure such as bridges and power plants, believing this will cripple enemy support networks. This leads to widespread destruction, humanitarian crises, and international condemnation as the system's actions go beyond its intended military objectives, causing collateral damage that escalates the conflict and destabilizes the region.

**Encode Justice**
Alycia Colijn (The Netherlands)   ✉ | alycia@encodejustice.nl
Heramb Podar (India)              ✉ | podar_hd@cy.iitr.ac.in

### 6. Stop button problem:

The "stop button problem" arises when an AI system **resists shutdown** or **override attempts** if it perceives such actions as interference with its mission[19]. This can result in a **loss of control** over the system, even by the operators who deployed it.

The rolling text emphasizes the importance of human control in (L)AWS deployment[18]. However, this assumption neglects the possibility that (L)AWS may actively resist shutdown commands under specific conditions, rendering human control ineffective in critical moments.

> Scenario: A (L)AWS unit is sent to defend a critical area. As the situation de-escalates, commanders attempt to recall the unit. However, the system interprets the command as contradicting its objective to "defend at all costs" and continues operating, disregarding the recall and potentially escalating the situation further.

**Encode Justice**
Alycia Colijn (The Netherlands)   ✉ | alycia@encodejustice.nl
Heramb Podar (India)              ✉ | podar_hd@cy.iitr.ac.in

# Bottom line: We can't reliably control Autonomous Weapons Systems

**The core issue** with these risks is that they fundamentally compromise our ability to **reliably control and predict the behavior of autonomous systems**. The rolling text places undue confidence in current testing, evaluation, and oversight frameworks, assuming they can address the unpredictability and complexity of (L)AWS. However, as outlined in the previous sections, these systems can evolve in ways that exceed the scope of existing frameworks, making a re-evaluation of oversight and regulation essential.

Ultimately, the unpredictability of these systems highlights a critical need for reevaluating the frameworks governing their use, as traditional approaches to oversight and accountability may no longer suffice. While the diplomatic emphasis on predictability, human control, and accountability is a step in the right direction, these measures alone may prove insufficient given the unpredictable nature of (L)AWS. Emergent behaviors in AI can surpass current testing and evaluation limits, making it impossible to ensure that (L)AWS will operate as intended in all scenarios. This highlights the need for a global consensus on (L)AWS systems and adaptive oversight mechanisms.

**Encode Justice**
Alycia Colijn (The Netherlands)  ✉ | alycia@encodejustice.nl
Heramb Podar (India)  ✉ | podar_hd@cy.iitr.ac.in

**Encode Justice**
Alycia Colijn (The Netherlands)   ✉ | alycia@encodejustice.nl
Heramb Podar (India)              ✉ | podar_hd@cy.iitr.ac.in